\documentstyle[aps,twocolumn]{revtex}

\begin{document}
\title{Attraction between pancakes vortices in the crossing lattices of layered
superconductors}
\author{A. Buzdin and I. Baladi\'{e}}
\address{Condensed Matter Theory Group, University Bordeaux I, CPMOH UMR-CNRS 5798,\\
33405 Talence Cedex, France}
\maketitle

\begin{abstract}
The intervortex interaction is investigated in very anisotropic layered
superconductors in tilted magnetic field. In such a case, the crossing
lattice of Abrikosov vortices (AVs) and Josephson vortices (JVs) appears.
The interaction between pancakes vortices (PVs), forming the AVs, and JVs
produces the deformation of the AV line. It is demonstrated that in the
result of this deformation a long range attraction between AVs is induced.
This phenomenon is responsible for the dense vortex chains formation. The
vortex structure in weak perpendicular magnetic field is the vortex chain
phase, while in higher field a more complicated mixed vortex chain-vortex
lattice phase emerges.
\end{abstract}


Vortex physics in layered superconductors occurred to be extremely rich and
interesting. In moderately anisotropic superconductors, a tilted magnetic
field leads to the formation of vortices inclined towards the
superconducting layers. The interaction between such tilted vortices happens
to be quite unusual.

In the plane defined by the vortex line direction and the ${\bf c}$-axis
(normal to the superconducting planes), the interaction between tilted
vortices is attractive at long distances. Such attractive intervortex
interaction is quite unexpected and leads to the formation of vortex chains,
where the intervortex distance is governed by the balance between the long
range attraction and the short range repulsion. In layered superconductors,
the existence of these vortex chains in tilted field has been predicted in 
\cite{Simonov}\ and \cite{Grishin}, and subsequently confirmed by the
decoration technique \cite{gavamel}, and the scanning-tunneling microscopy 
\cite{Hess} measurements, in YBa$_{2}$Cu$_{3}$O$_{7}$ and NbSe$_{2}$
crystals respectively. The systems studied in \cite{gavamel,Hess} are
characterized by a moderate anisotropy. The theoretical approach \cite
{Simonov,Grishin} is completely applicable to this case and gives a good
qualitative and quantitative description of this phenomenon. On the other
hand, in the much more anisotropic Bi$_{2}$Sr$_{2}$CaCu$_{2}$O$_{x}$ (BSCCO)
single crystals \cite{Bolle,Grigorieva}, a more complicated mixed vortex
chain-vortex lattice phase has been observed. As it has been demonstrated 
\cite{Bulaevski92}, in strongly anisotropic layered superconductors, in a
tilted magnetic field, a crossing lattice of Abrikosov and Josephson
vortices (JVs), a ``combined lattice'', must exist. The AV is in fact a line
of pancake vortices (PVs) \cite{Feinberg90,Clem,efetov,Artemenko}
interacting with JVs. Following \cite{koshelev}, the perpendicular vortex
line formed by the PVs is deformed and attracted by JVs, so the JVs stacks
accumulate additional PVs, creating vortex row with enhanced density \cite
{Bolle,Grigorieva}. This scenario has been proposed in \cite{koshelev} to
explain the mixed chain-lattice state formation. Very recently, using
scanning Hall probe microscopy, the detailed studies of vortex chains in
BSCCO have been performed and revealed the stability of the dense vortex
chains state even in the absence of lattice and in very weak perpendicular
magnetic field \cite{Bending}.

In the present work, we demonstrate that the formation of such dense vortex
chain is due to the attraction between the deformed lines of pancakes
vortices. The deformation, responsible for a long range attraction, appears
due to the interaction with JVs. In the result, the mechanism of vortex
chains formation, in tilted field, in strongly anisotropic superconductors,
appears to be quite similar to the case of moderately anisotropic
superconductors \cite{Simonov,Grishin}.

Further on, keeping in mind BSCCO, we will consider layered superconductors
with high anisotropy ratio $\gamma =\lambda _{c}/\lambda _{ab}\sim 200-500$,
where $\lambda _{c}$ is the penetration depth for currents along ${\bf c}$%
-axis (perpendicular to the layers), and $\lambda _{ab}$ is the penetration
depth for currents in the $ab$ plane (parallel to the layers). The in-plane
field $B_{x}=B\cos \theta $ penetrates inside the superconductor in the form
of JVs, while the perpendicular field $B_{z}=B\sin \theta $ creates the PVs
which interact with JVs via the Josephson coupling \cite
{koshelev,BulaevskiiMaley}. We consider the case of a very weak coupling of
the layers when the Josephson's core radius $\lambda _{J}=\gamma s$ ($s$ is
the interlayer spacing), is larger than an in-plane penetration depth, i.e. $%
\lambda _{J}>\lambda _{ab}.$ In \cite{BulaevskiiMaley}, the limit of high
field oriented near the $\left( {\bf a},{\bf b}\right) $ plane $B\gg
H_{0}=\phi _{0}/\gamma s^{2}$ was considered and it was demonstrated that a
zigzag displacement of PVs along ${\bf x}$-axis is produced, see Fig. $1$.
The authors of \cite{BulaevskiiMaley} analyzed the case of the dense vortex
lattice formed by PVs, but their method to treat the JVs and PV interaction
permits us to calculate the shape of the single zigzag line of PVs too.

In the limit of weak Josephson coupling, the interaction which stabilizes
the straight PVs line is mainly of electromagnetic origin. Then, we may use
the general expression for the energy of an arbitrary configuration of
pancakes in the framework of the electromagnetic model \cite{Feinberg90} to
calculate the energy increase due to the line deformation

\begin{eqnarray}
E_{em} &=&\frac{s}{8\pi \lambda _{_{ab}}^{2}}\sum_{n}\int \frac{d^{2}{\bf k}%
}{\left( 2\pi \right) ^{2}}\left| {\bf \Phi }_{n}({\bf k})\right| ^{2}-{\bf %
\Phi }_{n}(-{\bf k})  \nonumber \\
&&\times \sum_{m}{\bf \Phi }_{m}({\bf k})\frac{\sinh (kd)}{2\lambda _{eff}k}%
\frac{(G_{k}-\sqrt{G_{k}^{2}-1})^{\left| n-m\right| }}{\sqrt{G_{k}^{2}-1}},
\label{Energie générale}
\end{eqnarray}
where $\lambda _{eff}=\frac{\lambda _{_{ab}}^{2}}{s},${\bf \ }${\bf \Phi }%
_{n}({\bf k})$ is the Fourier transform of the total London vector of the $%
n^{th}$ layer ${\bf \Phi }_{n}(\overline{{\bf r}})=\sum {\bf \Phi }(%
\overline{{\bf r}}-{\bf r}_{pancake,n})$ (the sum is over all the pancakes
of the $n^{th}$ layer), ${\bf \Phi }({\bf k})=i\phi _{0}\frac{\left( {\bf k}%
\times {\bf z}\right) }{k^{2}},$ the function $G_{k}=\cosh \left( ks\right) +%
\frac{\sinh \left( ks\right) }{2\lambda _{eff}k}.$ For the zigzag
deformation with an amplitude $u,$ we have ${\bf \Phi }_{2n}({\bf k})=e^{i%
{\bf uk}}{\bf \Phi }({\bf k})$ and ${\bf \Phi }_{2n+1}({\bf k})=e^{-i{\bf uk}%
}{\bf \Phi }({\bf k}),$ and the energy increase per one layer occurs to be

\begin{equation}
\delta E_{em}=s\left( \frac{\phi _{0}}{4\pi \lambda _{_{ab}}}\right) ^{2}%
\frac{u^{2}}{\lambda _{_{ab}}^{2}}\ln \left( \frac{\lambda _{_{ab}}}{u}%
\right) .  \label{deform-energy}
\end{equation}
On the other hand, the gain of the Josephson energy due to PVs
displacements, calculated following \cite{BulaevskiiMaley} is $\delta
E_{J}=-u\left( \frac{\phi _{0}}{4\pi \lambda _{_{ab}}}\right) ^{2}\frac{%
2\phi _{0}}{\pi \gamma ^{2}s^{2}B_{x}}.$ Minimization of the total
deformation energy $\delta E_{em}+\delta E_{J}$ with respect to $u$ gives
its equilibrium value

\begin{equation}
u\sim \frac{\phi _{0}\lambda _{_{ab}}^{2}}{\gamma ^{2}s^{3}B_{x}}\frac{1}{%
ln\left( \frac{\gamma ^{2}s^{3}B_{x}}{\phi _{0}\lambda _{_{ab}}}\right) }=%
\frac{H_{0}}{B_{x}}\frac{\lambda _{_{ab}}^{2}}{\lambda _{J}}\frac{1}{%
ln\left( \frac{B_{x}\lambda _{J}}{H_{0}\lambda _{_{ab}}}\right) },
\label{ueq-1}
\end{equation}
and the limit $B_{x}\gg H_{0}$ corresponds to strongly overlapping Josephson
cores. Note that the amplitude of the zigzag modulation satisfies the
condition $u\ll \lambda _{_{ab}},$ and then it is a relatively small
deformation of the vortex line consisting of the PVs.

Now let us calculate the interaction of such two zigzag vortex lines at the
distance $x,$ both located in the $\left( {\bf x},{\bf z}\right) $ plane,
see Fig. 1. Those characteristic distances, which are of interest for us,
are much smaller than $\lambda _{c},$ and then in the calculation of the
interaction energy we may neglect the Josephson coupling and use the general
form of the energy in the pure electromagnetic limit $\left( \ref{Energie
générale}\right) $. In our case, the London vectors of the layers are ${\bf %
\Phi }_{2n}({\bf k})=e^{i{\bf uk}}{\bf \Phi }({\bf k})(1+e^{-i{\bf xk}})$
and ${\bf \Phi }_{2n+1}({\bf k})=e^{-i{\bf uk}}{\bf \Phi }({\bf k})(1+e^{-i%
{\bf xk}})$. Performing in $\left( \ref{Energie générale}\right) $ the
necessary summation and using for $G_{k}$ its expansion for $ks\ll 1$ (which
is perfectly justifiable when the distances of interest are larger than the
interlayer distance $s$), we finally obtain the following expression for the
interaction energy per one layer

\begin{equation}
E_{int}(x)=\frac{s\Phi _{0}^{2}}{2\pi \lambda _{_{ab}}^{2}}\int \frac{d^{2}%
\overrightarrow{k}}{\left( 2\pi k\right) ^{2}}\left[ 
\begin{array}{c}
\cos \left( xk_{x}\right) -\frac{\cos \left( xk_{x}\right) }{2\left(
1+k^{2}\lambda _{_{ab}}^{2}\right) } \\ 
-\frac{\cos \left[ k_{x}(x-2u)\right] }{4\left( 1+k^{2}\lambda
_{_{ab}}^{2}\right) }-\frac{\cos \left[ k_{x}(x+2u)\right] }{4\left(
1+k^{2}\lambda _{_{ab}}^{2}\right) }
\end{array}
\right] .  \label{Eint}
\end{equation}
Performing firstly the integration over $k_{y}$ and then over $k_{x},$ we
may present the interaction energy as

\begin{equation}
E_{int}(x)=s\left( \frac{\Phi _{0}}{4\pi \lambda _{_{ab}}}\right) ^{2}\left[ 
\begin{array}{c}
K_{0}\left( \frac{x}{\lambda _{_{ab}}}\right) +\frac{1}{2}K_{0}\left( \frac{%
x+2u}{\lambda _{_{ab}}}\right) \\ 
+\frac{1}{2}K_{0}\left( \frac{x-2u}{\lambda _{_{ab}}}\right) +ln\left( \frac{%
x^{2}-(2u)^{2}}{x^{2}}\right)
\end{array}
\right] ,  \label{Eint1}
\end{equation}
where $K_{0}$ is the modified Bessel function of zero order. At long
distances $x\gg \lambda _{_{ab}},$ the Bessel function $K_{0}\left( \frac{x}{%
\lambda _{_{ab}}}\right) $ decays exponentially, and the leading
contribution comes from the last term in $\left( \ref{Eint1}\right) $, which
gives $E_{int}(x)\approx -s\left( \frac{\Phi _{0}}{2\pi \lambda _{_{ab}}}%
\right) ^{2}\frac{u^{2}}{x^{2}}$, i.e. at long distances the net interaction
between the zigzag PVs lines is an attraction! At short distances, the
interaction is repulsive, as usual $E_{int}(x)\approx -s\left( \frac{\Phi
_{0}}{2\pi \lambda _{_{ab}}}\right) ^{2}ln\left( \frac{\lambda _{_{ab}}}{x}%
\right) .$ The overall behavior of the interaction energy is presented in
Fig. $2$ for different deformation parameters $\varepsilon =\frac{u}{\lambda
_{_{ab}}}\sim \frac{H_{0}}{B_{x}}\frac{\lambda _{_{ab}}}{\lambda _{J}}$. The
minimum energy may be easily found for small deformation parameter\ $%
\varepsilon $ using the asymptotic of the Bessel function $K_{0}\left(
z\right) \approx \sqrt{\frac{\pi }{2z}}exp(-z)$.\ With logarithmic accuracy
the minimum realizes at 
\begin{equation}
x_{min}\approx 2.5\lambda _{_{ab}}ln\left( \frac{3}{\varepsilon }\right)
\label{xmin}
\end{equation}
and 
\begin{equation}
E_{int}(x_{min})\approx -s\left( \frac{\Phi _{0}}{4\pi \lambda _{_{ab}}}%
\right) ^{2}\frac{\varepsilon ^{2}}{ln\left( \frac{1}{\varepsilon }\right) }.
\label{Emin}
\end{equation}
This result leads to the important physical conclusion that in the presence
of Josephson vortices, the AVs, due to the long range attractive
interaction, will form chains. The equilibrium distance between the vortices
in the chain is given, with logarithmic accuracy, by the same expression $%
\left( \ref{xmin}\right) $ as in the two vortices case. In fact, the energy
of the AV in chain is lower than the energy of a solitary vortex, and with
the increase of $B_{z}$ the vortices will start to penetrate in form of
chains. Further increase of $B_{z}$ will increase the number of chains
without changing in first approximation the distance between vortices in
chains. Only when the distance between the chains reaches the value of the
order of $\lambda _{_{ab}},$ the formation of the usual Abrikosov lattice
will occur. In the low field regime $B_{z}\ll H_{c1}^{c}=\frac{\phi _{0}}{%
4\pi \lambda _{ab}^{2}}\ln \left( \frac{\lambda _{_{ab}}}{\xi }\right) ,$
the well defined dense vortex chains must be present.

Now let us discuss the influence of the parallel field on the vortex chain
state. In the high field limit $B_{x}\gg H_{0},$ the increase of the
parallel field decreases the deformation parameter, see $\left( \ref{ueq-1}%
\right) $, and thus, following $\left( \ref{xmin}\right) $, will slightly
increase the intervortex distance inside the chain. A more important\
effect, actually, is the strong decrease of the potential dip in the
intervortex interaction energy $\left( \ref{Emin}\right) $ which will result
in the melting of the vortex chain. The vortex chain occurs to be more
stable at lower parallel field. However, the above approach is valid for $%
B_{x}>H_{0},$ and then the maximum stability of the chain is expected at $%
B_{x}\sim H_{0}$. At lower parallel field, the distance between Josephson
vortices along ${\bf z}$-axis becomes much larger than $2s,$ as well as the
distance between deformed parts of the PVs line. Evidently, the attraction
between AVs, due to its deformation, will weaken, and such a case needs a
special analysis which is presented below.

Let us consider the limit of weak parallel field, when the Josephson
vortices are well separated and the distance $D$ between them along ${\bf z}$%
-axis strongly exceeds the interlayer distance $s$, i.e. $B_{x}\ll H_{0}$,
see Fig. $3$. As it has been demonstrated in \cite{koshelev}, if the JV is
located between the layers $0$ and $1$, the pancake displacements $u_{n\text{
}}$on the $n^{th}$ layer in the case of the single AV is

\begin{equation}
u_{n\text{ }}\approx \frac{2C_{n}}{\left( n-\frac{1}{2}\right) }\frac{%
\lambda _{_{ab}}^{2}}{\lambda _{J}}\frac{1}{ln\left( \frac{\lambda _{J}}{%
\lambda _{_{ab}}}\right) },  \label{u-n}
\end{equation}
where $C_{n}$ is a numerical coefficient of order one. Note that as it may
be expected, this expression for $n=1$ is of the same order of magnitude as $%
\left( \ref{ueq-1}\right) $ at the limit of its applicability$\ $at $%
B_{x}\sim H_{0}.$

To treat this situation, it is convenient to add a fictitious pair of
pancake vortex and antivortex at the central line of the AV, see Fig. $3$.
Then, the obtained configuration will be equivalent to two straight
Abrikosov vortices and to two vortex-antivortex pairs at the distance $x$ in
each layer. Vortex and antivortex are separated by a distance $u_{n},$ and
we are coming to the problem of the interaction of such magnetic dipoles.
Firstly, note that the interaction between dipoles in the same layer is
attractive, and it may be directly calculated with the help of $\left( \ref
{Energie générale}\right) $:

\begin{equation}
E_{n,n}^{d}(x)\approx -\frac{s\Phi _{0}^{2}}{8\pi ^{2}\lambda _{_{ab}}^{2}}%
\frac{u_{n}^{2}}{x^{2}}.  \label{E-nn}
\end{equation}
It is much larger than the interaction between dipoles from different layers 
$n$ and $m,$ which may be attractive or repulsive. From $\left( \ref{Energie
générale}\right) $ it may be estimated as $\left| E_{n,m}^{d}(x)\right|
\approx \frac{s\Phi _{0}^{2}}{8\pi ^{2}\lambda _{_{ab}}^{2}}\frac{u_{n}u_{m}%
}{x^{2}}\frac{s}{\lambda _{_{ab}}}$\ for $s\left| n-m\right| \ll \lambda
_{_{ab}},$ i.e. containing the small additional factor $\frac{s}{\lambda
_{_{ab}}}\ll 1$. Moreover, for $s\left| n-m\right| >\lambda _{_{ab}},$ this
interaction decays exponentially. Taking into account that $u_{n}$ varies as 
$u_{n}\approx \frac{1}{\left( n-\frac{1}{2}\right) }$ near the JV, we may
finally conclude that the main contribution to the dipole interaction energy
is coming from the interaction in the same layer, and it may estimated (per
period $D$ of the modulation of PVs line along{\bf \ }${\bf z}$ axis) as

\begin{eqnarray}
E_{att}(x) &\approx &-\frac{s\Phi _{0}^{2}}{8\pi ^{2}\lambda _{_{ab}}^{2}}%
\frac{1}{x^{2}}\sum u_{n}^{2}  \nonumber \\
&\approx &-\frac{s\Phi _{0}^{2}}{8\pi ^{2}\lambda _{_{ab}}^{2}}\frac{%
u_{1}^{2}}{x^{2}}2\left( C_{1}^{2}+\frac{C_{2}^{2}}{3^{2}}+\frac{C_{3}^{2}}{%
5^{2}}+...\right)  \label{A} \\
&\approx &-\frac{s\Phi _{0}^{2}}{\lambda _{_{ab}}^{2}}\frac{u_{1}^{2}}{x^{2}}%
.
\end{eqnarray}
The main contribution to the repulsion energy is coming from the straight
AVs interaction, and per period $D$ at distances $x\gg \lambda _{_{ab}}$ is

\begin{equation}
E_{rep}(x)\approx \frac{D\Phi _{0}^{2}}{8\pi ^{2}\lambda _{_{ab}}^{2}}\sqrt{%
\frac{\pi \lambda _{_{ab}}}{2x}}exp\left( -\frac{x}{\lambda _{_{ab}}}\right)
.  \label{B}
\end{equation}
As in the case of the dense JVs lattice, the prevailing interaction between
AVs at long distances is an attraction, and it will lead to the vortex chain
creation. However, as the attraction is coming only from the PVs near JVs,
the relative strength of attraction is much smaller (other PVs contributing
only to the repulsion). In the result, the distance between AVs in the chain
may be estimated as

\begin{equation}
x_{min}\approx 2.5\lambda _{_{ab}}ln\left( \frac{1}{\widetilde{\varepsilon }}%
\right) ,  \label{C}
\end{equation}
where $\widetilde{\varepsilon }=\frac{u_{1}}{\lambda _{_{ab}}}\sqrt{\frac{s}{%
D}}\approx \frac{\lambda _{_{ab}}}{\lambda _{J}}\left( \frac{B_{x}}{H_{0}}%
\right) ^{1/4}$. And the energy gain due to the vortex chain formation per
pancake is

\begin{equation}
E_{int}\left( x_{min}\right) \approx -s\left( \frac{\Phi _{0}}{4\pi \lambda
_{_{ab}}}\right) ^{2}\frac{\widetilde{\varepsilon }^{2}}{ln\left( \frac{1}{%
\widetilde{\varepsilon }}\right) },  \label{D}
\end{equation}
i.e. it is decreasing with the decrease of $B_{x}$. Then, in the case $%
B_{x}\ll H_{0},$ the perpendicular vortices also appear as vortex chains,
and when the perpendicular field increases it will lead firstly to the
increase of the number of these chains, each chain located at JVs. The
distance between the chains will be an integer number of the distance
between JVs along ${\bf y}$-axis. Finally, when all JVs will contain chains,
with equilibrium distance $x_{\min }$ between AVs, further increase of $%
B_{z} $ will lead to the appearance of additional vortices in chains,
decreasing the intervortex distance below $x_{\min }$. In such a case, the
neighboring vortices repeal each other in the chain, but the repulsion
energy is compensated by the gain of energy due to the trapping of the AVs
by the JVs. However with further increase of $B_{z},$ the repulsion energy
will overcome the trapping energy and the formation of the usual Abrikosov
lattice will start. Namely this case corresponds to a mixed vortex
chain-vortex lattice observed in \cite{Bolle,Grigorieva}. It is interesting
that in low perpendicular field $B_{z}\ll H_{c1}^{c}$ and in the limit $%
B_{x}<H_{0}$ the mixed vortex chain-vortex lattice may transform into a
purely chain state with the increase of $B_{x}$. Indeed, the number of JVs
increases, permitting therefore to accommodate all AVs. This type of
behavior could be responsible for peculiar results on vortex lattice melting
in the presence of JVs obtained in \cite{Mirkovic}.

Comparing the results for low ($B_{x}\ll H_{0}$) and high ($B_{x}\gg H_{0}$)
parallel field limits, we see that the distance between AVs in chain is
always around $\lambda _{_{ab}}$ and slightly varies with $B_{x}$. On the
other hand, the energy of vortex coupling in a chain is maximal for $%
B_{x}\sim H_{0},$ and then they are more stable at this condition. At low or
high $B_{x}$ limits, we may expect the melting of vortex lines.

Note also that the vortex chain state could also reveal an anisotropy of the
critical current, the current along $y$ axis will provoke the motion of the
chain as a whole, while the current along $x$ axis will provoke the vortex
detachment from the chain.

In conclusion, we predict a new qualitative effect, the long range
attraction between AVs in the crossing vortex structure appearing in highly
anisotropic layered superconductors. Such an effect is somewhat reminiscent
to the tilted vortex attraction in superconductors with moderate anisotropy 
\cite{Simonov,Grishin}. If we consider the line between neighboring layers
with $\pi $ phase difference as the center of a JV, the PVs displacement
disrupts this line at the distance $2u$. So we may consider the resulting
structure as a tilted vortex line made from PVs and JVs parts (see the wavy
line in Fig. $3$). However the very important difference, with the straight
tilted vortices, is that in the highly anisotropic case the vortex
attraction is completely controlled by the parallel component of the
magnetic field only.

\begin{center}
{\large ACKNOWLEDGMENTS}
\end{center}

We thank S. Bending and J. Mirkovic for useful comments and correspondence.
We are grateful to M. Daumens and C. Meyers for stimulating discussions.
This work was supported by the ACI ''Supra-nanom\'{e}trique'' and ESF
''Vortex'' Programme.

\begin{center}
{\Large {\bf Figure Captions}}
\end{center}

FIG. $1$. Zigzag deformation of the PVs stacks in high parallel magnetic
field $\left( B_{x}\gg H_{0}\right) ,$ directed along the ${\bf x}$-axis. $u$
is the amplitude of deformation and the centers of PVs stacks are at the
distance $x$. JVs are presented by dashed lines.

FIG. $2$. The energy of interaction between two zigzag vortices as\ the
function of the distance between them for different deformation parameter $%
\varepsilon =\frac{u}{\lambda _{ab}}$ values.

FIG. $3$. Schematic picture of deformed AV lines in case of relatively weak
parallel magnetic field $\left( B_{x}\ll H_{0}\right) .$ The distance
between the centers of AV lines is $x$. The JVs are presented by dashed
lines. The period$\ D$ of JV's lattice along the ${\bf z}$-axis is much
larger than the interlayer distance $s$. The PVs separation breaks the JV
line, these parts are presented by cross, and the resulting ''tilted vortex
line'' is depicted by a wavy line. On the upper layer, the fictitious
vortex-antivortex pair is shown at the center of the AV. This procedure
restores a straight AV line with additional vortex dipole presented as a
shaded region.

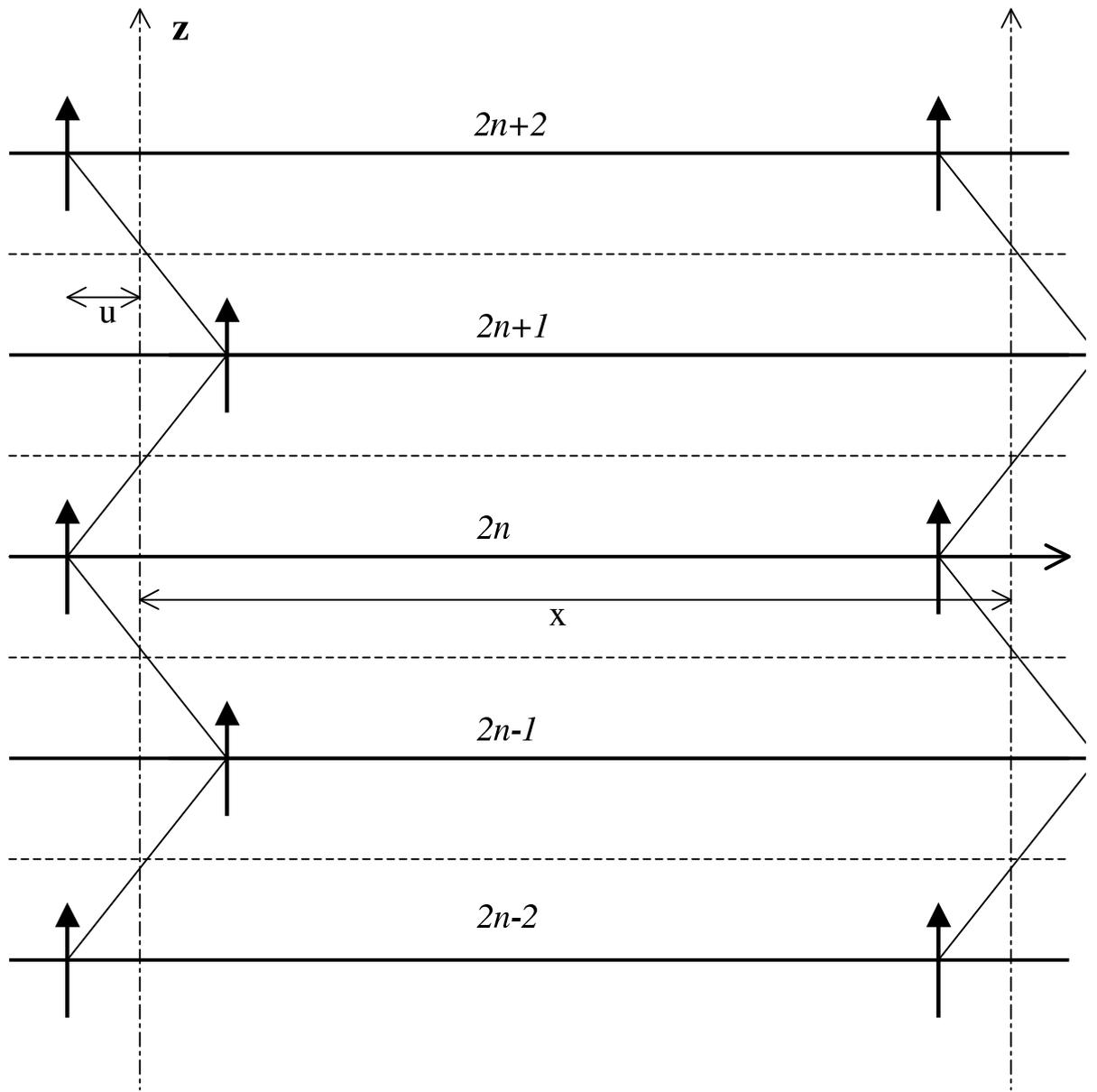
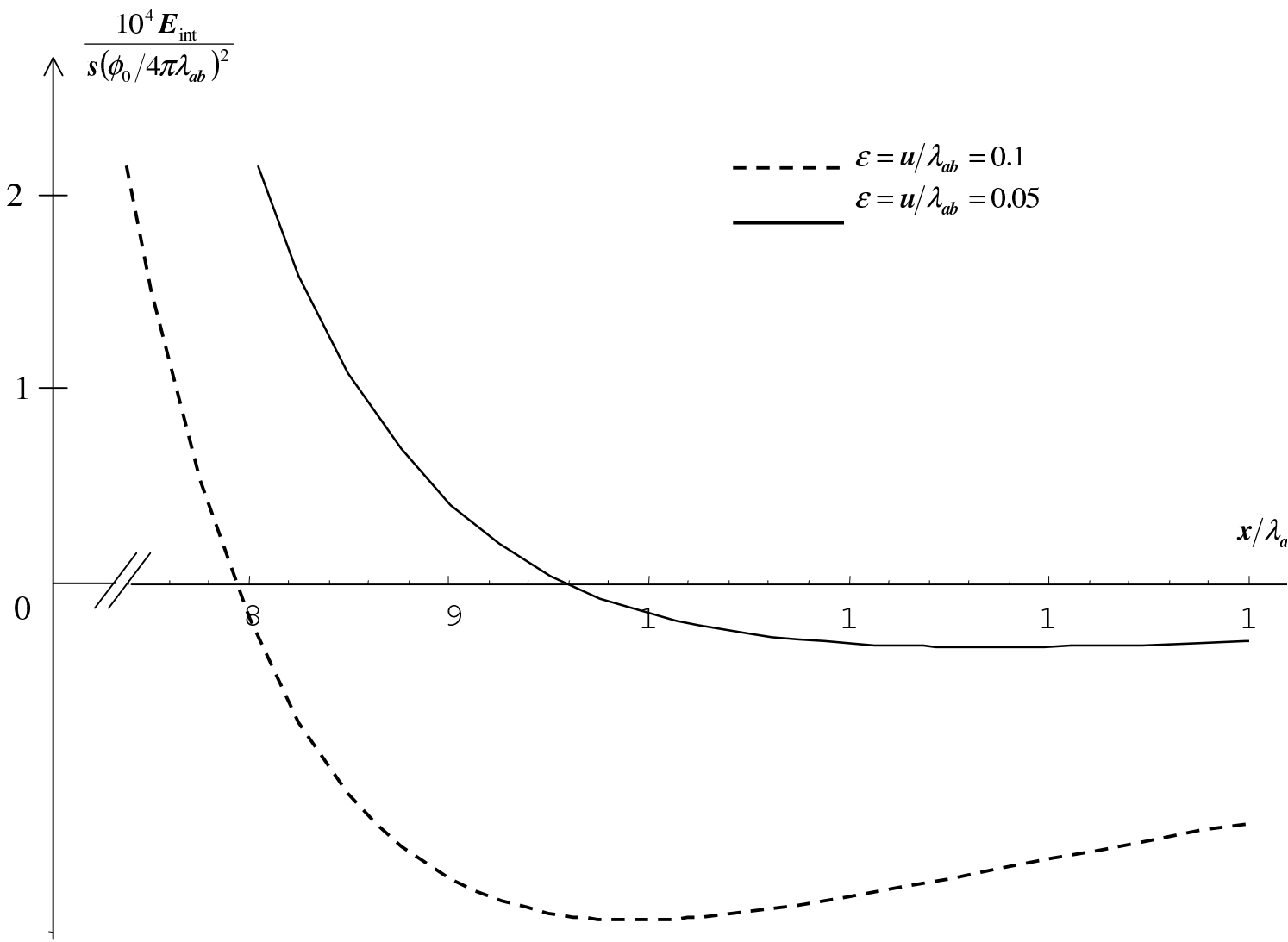
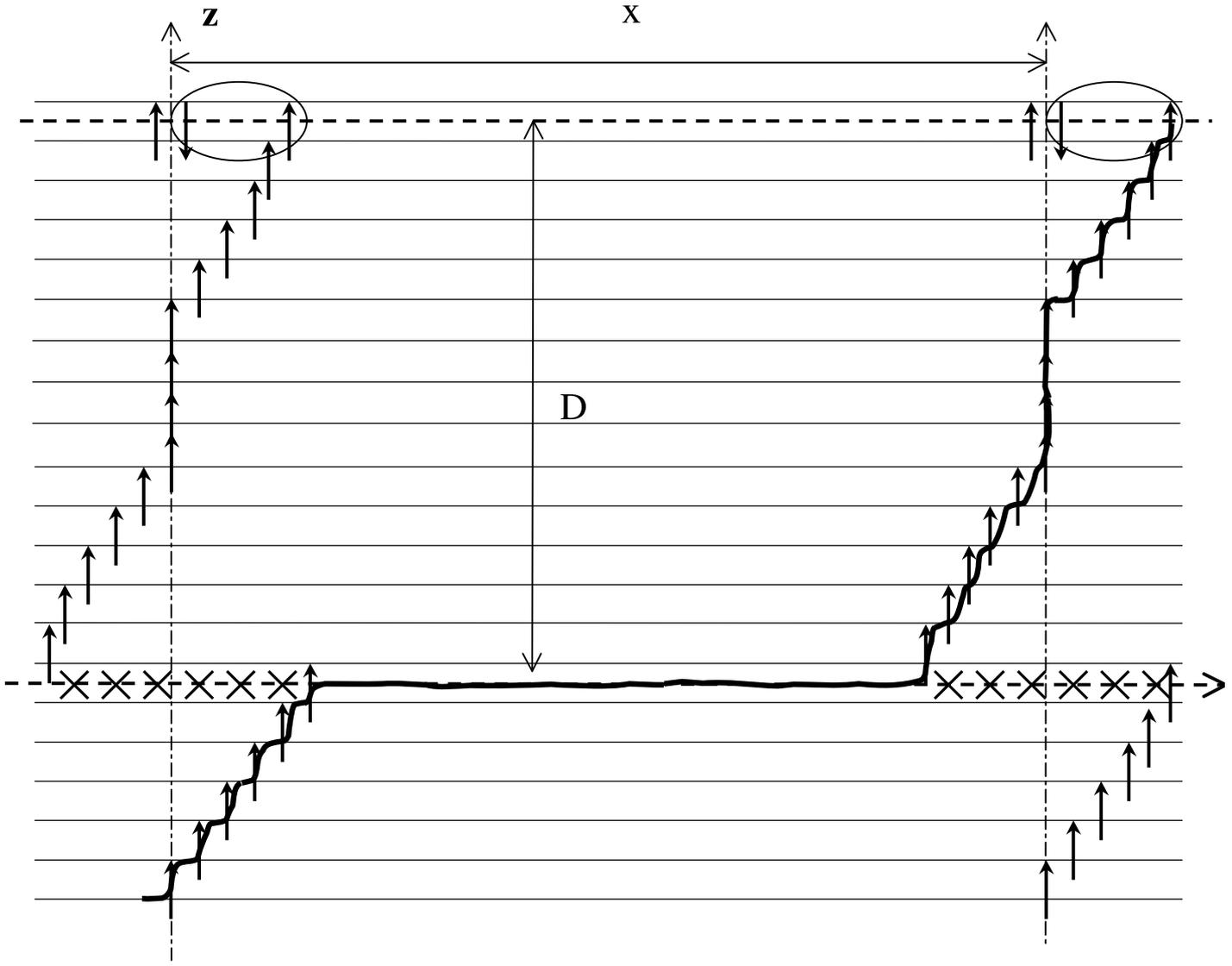
\end{document}